\documentclass{article}
\usepackage{bbm}
\usepackage{anysize}
\marginsize{3cm}{3cm}{4cm}{4cm}
\usepackage{makeidx}
\usepackage{amssymb}
\usepackage{amsmath}
\usepackage{mathrsfs}
\usepackage[Large]{subfigure}
\usepackage{graphicx}
\usepackage{lscape}
\usepackage{verbatim}
\usepackage{enumitem}
\usepackage{rotating}
\usepackage{tikz}

\newtheorem{theorem}{Theorem}

\newtheorem{definition}[theorem]{Definition}

\newtheorem{proposition}[theorem]{Proposition}
\newtheorem{remark}[theorem]{Remark}

\makeindex
\begin{document}
\author{Stefano M. Iacus and Lorenzo Mercuri}
\title{Implementation of L\'evy CARMA model in \texttt{Yuima} package}
\maketitle
\abstract{The paper shows how to use the \textsf{R} package \texttt{yuima} available on CRAN for the simulation and the estimation of a general L\'evy Continuous Autoregressive Moving Average (CARMA) model. The flexibility of the package is due to the fact that the user is allowed to choose several parametric L\'evy distribution for the increments. Some numerical examples are given in order to explain the main classes and the corresponding methods implemented in \texttt{yuima} package for the CARMA model.
}
\tableofcontents

%

\section{Introduction}

%


% ...
%
The Continuous Autoregressive Moving Average (CARMA) model driven by a standard Brownian Motion was first introduced in the literature by \cite{Doob1944} as a continuous counterpart of the discrete-time ARMA process and, recently, it has gained a rapid development in theory and practice. Indeed, in order to increase the level of appealing in different areas, the gaussianity assumption is relaxed and a CARMA model driven by a L\'evy process with finite second order moments has been introduced in \cite{Brockwell2001}. In this way the marginal distribution of the CARMA process is allowed to be asymmetric and heavy-tailed. For this reason the CARMA model is widely applied in the financial literature.
\newline For example, \cite{Barndorff2001} used a L\'evy CAR(1) (or Ornstein Uhlenbeck) process for building a stochastic volatility model while \cite{Todorov2006} and \cite{Todorov2011} applied the L\'evy CARMA(2,1)  for modeling the volatility of the Deutsche Mark/US Dollar daily exchange rate. Moreover \cite{Brockwell2005} proposed the fractionally integrated CARMA model in order to capture the long range dependence usually observed in financial time series.

The interest on the CARMA model is manifold since it can be used to model directly some given time series but it is also a main block for the construction of a more general process like the COGARCH(p,q) as in \cite{Brockwell2006}. 

The aim of this work is to develop in the \texttt{yuima} package a complete computational scheme for the simulation and the estimation of a general L\'evy CARMA model. Based on our knowledge, the \textsf{R} packages available on CRAN deal only with CARMA(p,q) models driven by a standard Brownian Motion \cite{ctarma} or  Gaussian CAR(p) models \cite{cts}. 
\newline For example the \texttt{ctarma} package developed by \cite{Tomasson2011} is an useful package for the simulation and the estimation of a CARMA(p,q) model driven by a Brownian Motion. Another package for continuous Autoregressive model is  the \texttt{cts} developed by \cite{cts} which deals with a modified version of the CAR(p) model named CZAR(p) by \cite{Tunnicliffe2004}. 
\newline Since the CAR(p) model is a special case of a CARMA(p,q), the \texttt{ctarma} package
is a valid benchmark for the functions implemented in the \texttt{yuima} package and for this reason a direct comparison is given in this paper where a Gaussian CARMA(p,q) model is considered.
Moreover, in the \texttt{yuima} package, once the estimation of the coefficients is done, it is possible to recover the underlying L\'evy process from the observed data using the methodology in \cite{BrockwellDavisYang2011} and extended to the multivariate CARMA(p,q) by \cite{BrockwellSchlemm2013}. In this way we are able to simulate trajectories of a
CARMA model without an explicit assumption on the distribution at time one of the underlying L\'evy process.
\newline The outline of this paper is the following. In Sect. \ref{TheoryCarma} we review the main results about the CARMA(p,q) process. In particular we focus the attention on the condition for the existence of the second order stationary solution of the CARMA process. In Sect. \ref{TwoStepEst} we explain the estimation procedure implemented in the \texttt{yuima} package if the data are observed in equally space-time intervals.
In Sect. \ref{ImplemCarmaInYuima} we describe the main classes and corresponding methods available in the \texttt{yuima} package for a CARMA model. We show how to use them for simulation and estimation of a Gaussian CARMA model and we conduct a comparison with the methods availables in the \texttt{ctarma} package. In Sect. \ref{NumEx} we present some numerical examples about the simulation and the estimation of  L\'evy CARMA models.  

\section{Continuous ARMA Models driven by a L\'{e}vy process}
\label{TheoryCarma}
In this section we review the main features of a CARMA(p,q) model driven by a L\'evy process introduced in \cite{Brockwell2001}

\begin{definition}
Let $p,\  q$ non-negative integers such that $p > q \geq 0$.  The CARMA(p,q) process is defined as:
\begin{equation}
a(D)Y_{t}=b(D)D L_{t}
\label{defCARMA1}
\end{equation}
$D$ is the differentation operator with respect to $t$ while $a\left(\cdot\right)$ and $b\left(\cdot\right)$ are two polynomials:
\begin{equation*}
a\left(z\right)=z^{p}+a_{1}z^{p-1}+\cdots+a_{p}
\end{equation*}
\begin{equation*}
b\left(z\right)=b_{0}+b_{1}z^{1}+\cdots+b_{p-1}z^{p-1}
\end{equation*}
where  $a_{1};\cdots;a_{p}$ and $b_{0},\cdots,b_{p-1}$ are coefficients such that $b_{q}\neq 0$ and $b_{j}=0$ $\forall j>q$.
\end{definition}
Since the higher order derivatives of a L\'evy process are not well defined we use the state space representation of a CARMA(p,q) model.
\begin{equation}
Y_{t}= \mathbf{b}^{\intercal}X_{t}
\label{carmaV}
\end{equation}
where $X_{t}$ is a vector process of dimension $p$ satisfying the following system of stochastic differential equations:
\begin{equation}
\mbox{d}X_{t}=AX_{t}\mbox{d}t + \mathbf{e}\mbox{d}L_{t}
\label{state-space-repr}
\end{equation}
where the $p \times p$ matrix $A$ is defined as:
\begin{equation*}
A \ = \ \left[
\begin{array}{ccccc}
 0 & 1 & 0 & \ldots & 0 \\
 0 & 0 & 1 & \ldots & 0 \\
\vdots & \vdots & \vdots & \ldots & \vdots \\
0 & 0 & 0 & \ldots & 1 \\
-a_{p} & -a_{p-1} & -a_{p-2} & \ldots & -a_{1} \\ 
\end{array}
\right]
\end{equation*}
\newline The $ p \times 1$ vectors $\mathbf{e}$ and $\mathbf{b}$ are respectively: 
\begin{equation*}
\mathbf{e}  = \left[ 0,\ldots,0,1\right]^{\intercal}
\end{equation*}
\begin{equation*}
\mathbf{b}  = \left[ b_{0},\ldots,b_{p-1}\right]^{\intercal}.
\end{equation*}
Given the initial condition on $X_{s}$, the solution of equation \eqref{state-space-repr} is:
\begin{equation}
X_{t}=e^{A\left(t-s\right)}X_{s}+\int_{s}^{t}e^{A\left(t-u\right)}\mathbf{e}\mbox{d}L_{u}, \ \ \forall t > s.
\label{expl_sol}
\end{equation}
Where the matrix exponential $e^{A}$ is defined as a power series:
\begin{equation*}
e^{A}=\sum_{h=0}^{+\infty}\frac{1}{h!}A^{k}.
\end{equation*}
The following result, given in \cite{Brockwell2001}, provides the necessary and sufficient conditions for the existence of a stationary solution $X_{t}$ of system \eqref{state-space-repr} such that $X_{t}$ is independent of $\left\{L_{h}-L_{s}, \ h > s\right\} \ \forall t \in R$

\begin{proposition}
\label{propsecord}
The process $X_{t}$ of system \eqref{state-space-repr} has a covariance stationary solution if and only if the real part of the eigenvalues $\lambda_{1}, \ \dots, \ \lambda_{p}$ of matrix $A$ are negative, i.e.
\begin{equation*}
\operatorname{Re} \left(\lambda_{i}\right)<0, \ \ i=1, \ \dots, \ p.
\end{equation*}
The solution can be written as:
\begin{equation}
X_{t}=\int_{-\infty}^{t}e^{A\left(t-u\right)}\mathbf{e}\mbox{d}L_{u}\buildrel d \over = \int_{0}^{+\infty}e^{Au}\mathbf{e}\mbox{d}L_{u}
\label{DebStatSol}
\end{equation}
and the associated first and second moments are:
\begin{equation*}
E\left[X_{t}\right]=\frac{\mu}{a_{p}}\mathbf{e}
\end{equation*}
\begin{equation*}
Cov\left[X_{t+h};X_{t}\right]=\sigma^{2}e^{Ah}\int^{+\infty}_{0}e^{Au}\mathbf{e}\mathbf{e}^\intercal e^{A^\intercal u}\mbox{d}u \ \ for \ \ h \geq 0. 
\end{equation*}
where $\mu=E\left[L_{1}\right]$ and $\sigma^{2}=Var\left[L_{1}\right]$
\end{proposition}

\begin{remark}
We observe that matrix $A$ can be diagonalized as follows:
\begin{equation*}
A=R \Lambda R^{-1}.
\end{equation*}
$\Lambda$ is a matrix whose elements along the diagonal are the eigenvalues of $A$ and the other elements are zero.
\begin{equation*}
\Lambda= diag(\lambda_{j}) \ \ \ j=1, \ldots,p. 
\end{equation*}
The columns of $R$ are the eigenvectors of $A$ which are obtained easily from the eigenvalues:
\begin{equation*}
R_{,j}=\left[1,\lambda_{j},\lambda^{2}_{j},\ldots,\lambda_{j}^{p-1}\right]^{\intercal}, \ \ \ j=1,\ldots,p.
\end{equation*}
The necessary and sufficient condition for the diagonalization of $A$ is that the eigenvalues $\lambda_{j}, \ \ \ j=1,\ldots,p$ are distincts.
\end{remark}
Using equation \eqref{DebStatSol}, the solution of CARMA process $Y_{t}$ has the following form:
\begin{equation*}
Y_{t}=\mathbf{b}^{\intercal}X_{t}=\int_{-\infty}^{+\infty}g\left(t-u\right)\mbox{d}L_{u}.
\end{equation*}
where $g\left(t\right)=\mathbf{b}^{\intercal}e^{At} \mathbf{e} \mathbbm{1}_{\{ \left[ 0, +\infty \right) \}}\left(t\right)$ is the Kernel of the CARMA process $\left\{Y_{t}\right\}$ and $\mathbbm{1}_{\{A \}}\left(x\right)$ is the indicator function defined as:
\begin{equation*}
\mathbbm{1}_{\{A \}}\left(x\right):=\left\{
\begin{array}{lll}
1 & & x \in A\\
0 & & x \notin A \\
\end{array}
\right.
\end{equation*}. 
\begin{proposition}
Under the assumptions that the eigenvalues of matrix $A$ are distinct and $\operatorname{Re} \left(\lambda_{i}\right)<0$ for $i=1,\ldots,p.$ the CARMA(p,q) process can be obtained as a sum of dependent CAR(1) processes:
\begin{equation}
Y_{t}=\sum_{r=1}^{p}Y_{r,t}
\label{repr1}
\end{equation}
where
\begin{equation}
Y_{r,t}=\int_{-\infty}^{t}\alpha_{r}e^{\lambda_{r}\left(t-u\right)}\mbox{d}L_{u},
\end{equation}
\begin{equation}
\alpha_{r}=\frac{b\left(\lambda_{r}\right)}{a'\left(\lambda_{r}\right)}, \ \ \ r=1,\ldots,p.
\label{alpha_r}
\end{equation}
and $a'(z)$ is the first derivative of the polynomial $a\left(z\right)$.
\end{proposition}
In particular, the vector $\tilde{Y_{t}}=\left[Y_{1,t},\ldots,Y_{p,t}\right]$, whose elements are the CAR(1) processes necessary in the representation \eqref{repr1}, can be obtained as:
\begin{equation}
\tilde{Y}_{t}= \tilde{\Lambda} R^{-1}X_{t}
\label{canonform}
\end{equation}
Where $\tilde{\Lambda}$ is a diagonal matrix defined as:
\begin{equation*}
\tilde{\Lambda}=\operatorname{diag}\left[ b\left(\lambda_{i}\right)\right]_{i=1}^{p}.
\end{equation*}
This is the canonical representation of CARMA process, the vector $\tilde{Y_{t}}$ is the canonical state vector and it will be useful for recovering the increments of the underlying noise.


%

\section{Estimation of a CARMA(p,q) model in the \texttt{yuima} package}
\label{TwoStepEst}

%
%
%
%
%
%
%
%
%

In this Section we discuss the estimation procedure implemented in the \texttt{yuima} package for a CARMA model driven by a L\'evy process. From now on, we assume that the condition for canonical state representation (i.e. distinct eigenvalues for $A$ matrix whose real part is negative) is satisfied. As observed before, we consider a three step procedure:
\begin{enumerate}
\item Exploiting the state space representation, we estimate the CARMA parameters $ \mathbf{a}:=\left[ a_{1},\ldots,a_{p}\right]$ and $\mathbf{b}:=\left[ b_{0},\ldots,b_{q},b_{q+1}=0,\ldots, b_{p-1}=0\right]$ through the quasi-maximum likelihood estimation 
(see \cite{Schlemm2012} for univariate and multivariate cases).
 An alternative approach is based on the Least Square estimation 
 (see \cite{BrockwellDavisYang2011} for more details).Since the state space representation in system \eqref{state-space-repr} is based on the unbservable process $X_{t}$, we implement a Kalman Filter  procedure (see \cite{Tomasson2011} for a CARMA model driven by a brownian motion).
\item Once the CARMA parameters have been found, we recover the increments of the underlying L\'evy following the approach proposed in \cite{BrockwellDavisYang2011} as a generalization of the approach developed in \cite{BrockwellDavisYang2007} for the continuous autoregressive process. Recently the same approach has been applied to the multivariate case by \cite{BrockwellSchlemm2013}. 
\item In the last step, using the increments estimated in the previous step, we estimate the parameters of the L\'{e}vy measure. The likelihood function is computed by means of the Fourier Transform for all L\'evy increments assumption available in the \texttt{yuima} package.
\end{enumerate}
Following \cite{BrockwellDavisYang2011} we assume that the observations $Y_{1},\ldots,Y_{n},\ldots, Y_{T}$ are collected at equally spaced time instants $0, h, 2h, \dots, Nh$ where $N$ is the number of obsevations and $h$ is the step length. In this context, the time horizon $T$ is equal to $Nh$.
\newline In order to be more general as possible, the expectation and the variance of the L\'evy at time 1 are given by:
\begin{equation*}
E\left[ L_{1} \right]=\mu \ \ \ Var\left[ L_{1} \right]=\sigma^2.
\end{equation*}
Then we define a mean corrected process $Y^{\star}_{n}$ as:
\begin{equation}
Y^{\star}_{n}=Y_{n}-E\left[ Y_{n} \right].
\label{meancorrproc}
\end{equation}
Let $Y_{n}$ be a ergodic series, we estimate the expectation in \eqref{meancorrproc} using the sample mean. Finally, the process $Y^{\star}_{n}$ has again a state space representation:
\begin{eqnarray*}
Y^{\star}_{n}&=& \mathbf{b}^{\intercal}X^{\star}_{n}\\
X^{\star}_{n}&=& e^{Ah}X^{\star}_{n}+U_{n}\\
\end{eqnarray*}
Where $X^{\star}_{n}$ is a sample mean corrected state vector process of the $X_{t}$ in system \eqref{state-space-repr}. $e^{A}$ is the matrix exponential of $A$. $U_{n}$ is a sequence of i.i.d random vectors with zero mean and variance-covariance matrix:
\begin{equation}
Q=\int_{0}^{h}e^{Au}\mathbf{e}\mathbf{e}^{\intercal}e^{A^{\intercal}u}\mbox{d}u
\label{Qmatrix}
\end{equation}

\begin{remark}
If $Y_{t}$ is a CARMA process driven by a Brownian Motion then the sampled process $Y_{n}$ is a Gaussian ARMA process with i.i.d. noise for any step length $h$. For a second order L\'evy CARMA process, the driving noise is not necessarly i.i.d. but the sampled process $Y_{n}$ is still an ARMA process. In this case the process $Y^{\star}_{n}$ is a weak ARMA process and the distribution of the maximum likelihood estimators can be derived using the result in \cite{Francq1998}. 
\end{remark}

Before introducing the Kalman Filter algorithm (see \cite{Kalman1960} for more details) we need to compute the $Q$ matrix in \eqref{Qmatrix}. We start by evaluating the stationary unconditional variance-covariance matrix $Q_{\infty}$ satisfying the system of equations (see \cite{Tsai2000} for more details):
\begin{equation*}
AQ_{\infty}+Q_{\infty}A^{\intercal}=-\sigma^2 \mathbf{e}\mathbf{e}^{\intercal} 
\end{equation*}
then the matrix $Q$ is obtained using the following formula:
\begin{equation*}
Q=Q_{\infty}-e^{Au}Q_{\infty}e^{A^{\intercal}u}
\end{equation*}
Using this result, the Kalman Filter algorithm gives us a simple and analytical way for computing the likelihood function. The estimation procedure based on the Kalman Filter can be summarized into four steps: initialization, prediction, correction and construction of the log-likelihood function. 
\newline Before explaining the steps in the Kalman Filter algorithm we need to clarify the used notations. We define $X^{\star}_{n\left|n-1\right.}$ and $\Sigma_{n\left|n-1\right.}$ the prior estimates of the state variables process and the Variance-Covariance matrix of the error term $U_{n}$, i.e.
\begin{eqnarray*}
X^{\star}_{n\left|n-1\right.}&=&E\left[X^{\star}_{n}\right|\left.I_{n-1}\right]\\
\Sigma_{n\left|n-1\right.}&=&Var\left[U_{n}\right|\left.I_{n-1}\right]\\
\end{eqnarray*}
where the $\sigma-$algebra $I_{n-1}$ is generated by the observations of the process $Y^{\star}_{n}$ and by the estimates of the state space variables up to time $n$:
\begin{equation*}
I_{n-1}=\sigma\left(Y^{\star}_{n-1},\ldots,Y^{\star}_{0},X^{\star}_{n-1\left|n-1\right.},\ldots,X^{\star}_{0\left|0\right.}\right).
\end{equation*}
We denote with $X^{\star}_{n\left|n\right.}$ and $\Sigma_{n\left|n\right.}$ the posterior estimates for the mean corrected state process $X^{\star}$ and the Variance-Covariance matrix for the process $U_{n}$. In this case the estimates are obtained according to the augmented $\sigma-$algebra $I^{\star}_{n}$:
\begin{equation*}
I^{\star}_{n}=\sigma\left(\left\{I_{n-1}\right\}\cup\left\{Y^{\star}_{n},X^{\star}_{n\left|n-1\right.}\right\}\right).
\end{equation*}
\textbf{Initialization.}\newline
We initialize the state variable $X^{\star}_{n}$ at zero since, under the assumption that the eigenvalues of Matrix $A$ are distinct with negative real part, the unconditional mean $E\left[ X^{\star}_{n} \right]$ is equal to zero while the variance-covariance matrix is initialized at the uncoditional variance-covariance matrix $Q_{\infty}$. Finaly we set:
\begin{eqnarray*}
X^{\star}_{0\left|0\right.} &=& 0 = X^{\star}_{n-1\left|n-1\right.}\\
\Sigma_{0\left|0\right.} &=& Q_{\infty} =\Sigma_{n-1\left|n-1\right.}\\
\end{eqnarray*}
\textbf{Prediction.}\newline
We start by predicting the unobservable process $X^{\star}_{n\left|n-1\right.}$ and the variance-covariance matrix $\Sigma_{n\left|n-1\right.}$
\begin{eqnarray}
X^{\star}_{n\left|n-1\right.} &=& e^{Ah}X^{\star}_{n-1\left|n-1\right.} \nonumber\\
\Sigma_{n\left|n-1\right.} &=& e^{Ah}\Sigma_{n-1\left|n-1\right.}e^{A^{\intercal}h}+Q \nonumber\\
\label{init-step}
\end{eqnarray}
then we can forecast the observable process $Y_{n}$:
\begin{equation*}
Y^{\star}_{n\left|n-1\right.}=b^{\intercal}X^{\star}_{n\left|n-1\right.}.
\end{equation*}
We define the error term $u_{n}$ as:
\begin{equation*}
u_{n}=Y^{\star}_{n}-b^{\intercal}X^{\star}_{n\left|n-1\right.}
\end{equation*}
where $Y^{\star}_{n}$ is the observed mean corrected process defined in \eqref{meancorrproc}. Thus, the error term $u_{n}$ is normally distributed 
\begin{equation*}
u_{n} \sim N\left[ 0, \ \mathbf{b}^{\intercal}\Sigma_{n\left|n-1\right.}\mathbf{b} \right]
\end{equation*}
and we use the result to build the log-likelihood function.\smallskip \newline
\textbf{Correction} \newline
We need to update the state variable $X^{\star}$ and the variance-covariance matrix $\Sigma$ since we observe the realization of the process $Y^{\star}_{n}$.
\begin{eqnarray*}
X^{\star}_{n\left| n \right.} &=& X^{\star}_{n\left| n - 1 \right.}+K_{n}\left( Y^{\star}_{n} - Y^{\star}_{n \left| n-1 \right.} \right) \nonumber\\
\Sigma_{n\left|n\right.} &=& \Sigma_{n \left| n - 1 \right.}-K_{n}\mathbf{b}^{\intercal}\Sigma_{n \left| n - 1 \right.}
\end{eqnarray*}
where $K_{n}$ is the Kalman Gain Matrix and it is defined as:
\begin{equation*}
K_{n}=\Sigma_{n \left| n - 1 \right.}\mathbf{b}\left(\mathbf{b}^{\intercal}\Sigma_{n \left| n - 1 \right.}\mathbf{b}\right)^{-1}.
\end{equation*}
We use the updated state variable $X_{n\left| n \right.}$ and the variance-covariance matrix $\Sigma_{n\left|n\right.}$ as inputs in \eqref{init-step} and we repeat steps $1 \div 3$ until $n=N$.
\smallskip
\newline
\textbf{Construction of the log-likelihood function}
\newline
Once all error terms $\left\{u_{n}\right\}_{n=1}^{N}$ are obtained, we compute the log-likelihood function:
\begin{equation}
\mathscr{L}\left(\mathbf{a},\mathbf{b}\right)=-\frac{1}{2}\sum_{n=1}^{N}\ln\left(2 \pi \mathbf{b}^{\intercal}\Sigma_{n\left| n-1 \right.}\mathbf{b}\right)-\frac{1}{2}\sum_{n=1}^{N}\frac{u_{n}^{2}}{\mathbf{b}^{\intercal}\Sigma_{n\left|n-1\right.}\mathbf{b}}
\label{eq:loglik}
\end{equation}
and get the estimates for vectors $\mathbf{a}$ and $\mathbf{b}$ by maximizing the quantity in \eqref{eq:loglik}. In the \texttt{yuima} package constrained optimization is also available.\smallskip \newline
Once the estimates for vectors $\mathbf{a}$ and $\mathbf{b}$ are obtained, the next step is to retrieve the increments of underlying L\'evy process. It is worth to notice that the procedure for recovering the underlying L\'evy increments is a non-parametric approach since the knowledge of the distribution is not necessary at this stage while it becomes relevant in the last passage of the estimation procedure implemented in \texttt{yuima} package. \newline
Following \cite{BrockwellDavisYang2011}, the vector $\tilde{X}_{q,t}$ composed by the firsts $q-1$ components of the state process $X_{t}$ in \eqref{expl_sol}, can be written in terms of the observable process $Y_{t}$.
\begin{equation}
\mbox{d}\tilde{X}_{q,t} = B\tilde{X}_{q,t}\mbox{d}t+\mathbf{e_{q}}Y_{t}\mbox{d}t
\label{rapreseforXq}
\end{equation}
where the matrix $B$ is defined as:
\begin{equation*}
B \ = \ \left[
\begin{array}{ccccc}
 0 & 1 & 0 & \ldots & 0 \\
 0 & 0 & 1 & \ldots & 0 \\
\vdots & \vdots & \vdots & \ldots & \vdots \\
0 & 0 & 0 & \ldots & 1 \\
-b_{0} & -b_{1} & -b_{2} & \ldots & -b_{q-1} \\ 
\end{array}
\right]
\end{equation*}
and the vector $\mathbf{e_{q}}$ 
\begin{equation*}
\mathbf{e_{q}}  = \left[ 0,\ldots,0,1\right]^{\intercal}
\end{equation*}
The system of equations \eqref{rapreseforXq} has the explicit solution:
\begin{equation*}
\tilde{X}_{q,t} = e^{Bt}\tilde{X}_{q,0}+\int_{0}^{t}e^{B\left(t-u\right)}\mathbf{e_{q}}Y_{u}\mbox{d}u
\label{explicitSolXq}
\end{equation*}
The remaining $p-q$ components of $X_{t}$ are obtained by computing the higher order derivatives of the first component $X_{0,t}$ of the state vector $X_{t}$ with respect to time:
\begin{equation}
X_{j,t}=\frac{\partial^{j-1} X_{0,t}}{\left(\partial t \right)^{j-1}}, \ \ \ j=q, \dots, p-1.
\end{equation}
Using the canonical form of a CARMA process in \eqref{canonform}, we obtain the canonical state vector $\tilde{Y_{t}}$ and, following \cite{BrockwellDavisYang2007}, the underlying L\'evy can be expressed using one of the equation in the following system:
\begin{equation}
L_{t}=\frac{1}{\alpha_{r}}\left[\tilde{Y_{r,t}}-\tilde{Y_{r,0}}-\lambda_{r}\int_{0}^{t}Y_{r,u}\mbox{d}u\right] \ \ \ r = 1, \ldots,p.
\label{syst_L}
\end{equation}
where $\alpha_{r}$ is defined in equation \eqref{alpha_r} and $\lambda_{r}$ is the \textit{r}th eigenvalue of the matrix $A$. For estimation of the L\'evy, as suggested in \cite{BrockwellDavisYang2011}, we choose the condition in \eqref{syst_L} such that the corresponding $\lambda_{r}$ is the largest real eigenvalue.\smallskip \newline
Once the increments of the underlying L\'evy are obtained, in the \texttt{yuima} package, it is possible to estimate the parameters of the L\'evy measure. We refer to the yuima documentation (see \cite{yuimaPack} for  more details for the available L\'evy processes in \texttt{yuima} package. The estimation procedure in this phase is the maximum likelihood and the density is obtained by inverse Fourier Transform.

\section{Implementation of a CARMA(p,q) in the \texttt{Yuima} package}
\label{ImplemCarmaInYuima}
This Section is devoted to the description of the objects and methods available in the \texttt{yuima} package for defining a general CARMA model driven by a L\'evy process in the \textsf{R} statistical environment \cite{RdevelopTeam}. 

The \texttt{yuima} package \cite{yuimaPack} is a comprehensive framework  based on the S4 system of classes and methods (see \cite{Chambers1998} for a complete treatement of the S4 class system) which allows a description of stochastic differential equations with the following form:  
\begin{equation*}
\mbox{d}X_{t}=b\left(t, X_{t}\right)\mbox{d}t+\sigma\left(t,X_{t}\right)\mbox{d}W^{H}_{t}+c\left(t,X_{t}\right)\mbox{d}Z_{t}.
\end{equation*}
where $b\left(t, X_{t}\right)$, $\sigma\left(t,X_{t}\right)$ and $c\left(t,X_{t}\right)$ are coefficients defined by the user. $W^{H}_{t}$ is a fractional Brownian motion and $H$ is the Hurst index which default value is fixed to $\frac{1}{2}$ corresponding to the case of the standard Brownian motion (see \cite{BrousteIacus2013} for estimation of $H$ index in \texttt{yuima} package) and $Z_{t}$ is a pure L\'evy jump process (see \cite{Bertoin1998, Sato1999} for more details).

In this context, the mathematical description of a CARMA(p,q) process is done by the \texttt{yuima} constructor function \texttt{setCarma} that returns an object of class \texttt{yuima.carma}. Since the \texttt{yuima.carma-class} extends the \texttt{yuima.model-class} (see \cite{Brousteetal2013} for a complete description of an object of class \texttt{yuima.model}), it is possible to generate a sample path using the \texttt{simulate} method, estimate the parameters applying the \texttt{qmle} method and it is also available the utility \texttt{toLatex} that produce a \LaTeX code that returns the state space representation of the CARMA(p,q) model using the matrix notations.  The method \texttt{CarmaNoise} works only for object of class \texttt{yuima.carma} and allows to retrieve the increments of the underlying L\'evy following the approach described in Sect. \ref{TwoStepEst} once the vectors $\mathbf{a}$ and $\mathbf{b}$ are known. 

\subsection{The \texttt{yuima.carma-class}}

An object of the class \texttt{yuima.carma} contains all informations related to a general linear state space model that encompasses the CARMA model illustrated in Sect. \ref{TheoryCarma}.\newline The mathematical description of this general model is given by the following system of equations:
\begin{eqnarray}
Y_{t}&=&c_{0}+\sigma \left(\mathbf{b}^{\intercal} X_{t}\right)\nonumber\\
\mbox{d}X_{t}&=&A X_{t}\mbox{d}t+\mathbf{e}\left(\gamma_{0}+\gamma^{\intercal} X_{t} \right)\mbox{d}Z_{t}\nonumber\\
\label{gen-state-space-mod}
\end{eqnarray}
where $c_{0} \in \mathcal R  $ and $\sigma \in \left(0,+\infty \right)$ are location and scale parameters respectively. The vector $\mathbf{b} \in \mathcal R^{p}$ contains the moving average parameters $b_{0}, b_{1}, \ldots, b_{q}$ while the $A$ is a $p \times p$ matrix whose last row contains the autoregressive parameters $a_{1},\ldots,a_{p}$ and, as shown in Sect.  \ref{TheoryCarma}. It is defined as:
\begin{equation*}
A \ = \ \left[
\begin{array}{ccccc}
 0 & 1 & 0 & \ldots & 0 \\
 0 & 0 & 1 & \ldots & 0 \\
\vdots & \vdots & \vdots & \ldots & \vdots \\
0 & 0 & 0 & \ldots & 1 \\
-a_{p} & -a_{p-1} & -a_{p-2} & \ldots & -a_{1} \\ 
\end{array}
\right].
\end{equation*}
The $\gamma_{0} \in \mathcal R$ and the vector $\gamma:=\left[\gamma_{1},\ldots \gamma_{p}\right]$ are called linear parameters. The linear parameters $\left[\gamma_{0},\gamma_{1},\ldots \gamma_{p}\right]$   will play a central role for defining the COGARCH(p,q) model introduced in \cite{Brockwell2006} in the \texttt{yuima} package that will be one of the main objects of future developements.\newline
As noted previously, the \texttt{yuima.carma}  extends the \texttt{yuima.models} and all features in this class are inherited. In particular the structure of an object of class \texttt{yuima.carma} is composed by the slots listed below:
\begin{itemize}
    \item{\texttt{info} }{is an object of \texttt{carma.info-class} that describes the structure of the CARMA(p,q) model.}
    \item{\texttt{drift} }{is an \textsf{R} expression which specifies the drift 
   coefficient (a vector).}
    \item{\texttt{diffusion} }{is an \textsf{R} expression which specifies the diffusion 
     coefficient (a matrix).}
    \item{\texttt{hurst} }{is the Hurst parameter of the fractional Brownian 
	 motion. The default value $\frac{1}{2}$ corresponds of the standard Brownian process.}
    \item{\texttt{jump.coeff} }{is a vector of \texttt{expression}s for the jump 
	 component.}
    \item{\texttt{measure} }{indicates the measure of the L\'evy process.}
    \item{\texttt{measure.type} }{is a switch variable that indicates if the type of L\'evy measure specified in the slot \texttt{measure} belongs to the class of Compound Poisson processes.}
    \item{\texttt{state.variable}} {indicates a vector of names identifying the names used to 
     denote the state variable in the drift and diffusion specifications.}
    \item{\texttt{parameter}}{ is a short name for ``parameters'', is an 
	 object of class \texttt{model.parameter-class}. For more details see 
	 \texttt{yuima} documentation.}
    \item{\texttt{state.variable}}{ identifies the state variables in the \textsf{R} 
     expression.}
    \item{\texttt{jump.variable}}{ identifies the variable for the jump 
	 coefficient.}
    \item{\texttt{time.variable}}{is the name of the time variable.}
    \item{\texttt{noise.number}}{ denotes the number of sources of noise. 
	 Currently only for the Gaussian part.}
    \item{\texttt{equation.number}}{ is the dimension of the stochastic 
     differential equation.}
    \item{\texttt{dimension}}{ is the dimension of the parameter given in the slot 
     \texttt{parameter}.}
    \item{\texttt{solve.variable}}{ identifies the variable with respect to which 
	 the stochastic differential equation has to be solved.}
    \item{\texttt{xinit}}{ contains \textsf{R} expressions that are the initial conditions for the stochastic differential equations.}
    \item{\texttt{J.flag}}{ is for internal use only}.
\end{itemize}
It is worth to remark that, except for the slot \texttt{info}, the remainings are members of the \texttt{yuima.model-class}. Indeed the object of class \texttt{carma.info} in the slot \texttt{info} contains all informations about the CARMA model. It cannot be directly specified by the user but it is constructed by \texttt{setCarma} function that fills the following slots:
\begin{itemize}
    \item{\texttt{p}}{ is a integer number the indicates the dimension of autoregressive coefficients.}
    \item{\texttt{q}}{ is the dimension of moving average coefficients.}
    \item{\texttt{loc.par}}{ is the label of location coefficient.}
    \item{\texttt{scale.par}}{ indicates the Label of scale coefficient.}
    \item{\texttt{ar.par}}{ denotes the label of autoregressive coefficients.}
    \item{\texttt{ma.par}}{ is the label of moving average coefficients.}
    \item{\texttt{lin.par}}{ indicates the label of linear coefficients.}
    \item{\texttt{Carma.var}}{ denotes the label of the observed process.}
    \item{\texttt{Latent.var}}{ is the label of the state process.}
    \item{\texttt{XinExpr}}{ is a logical variable. If \texttt{XinExpr=FALSE}, the starting condition of \texttt{Latent.var} is zero otherwise each component of \texttt{Latent.var} has a parameter as a starting point.}
\end{itemize}

\subsection{CARMA model specification}
\label{CARMA1spec}

In this section we explain how to use the constructor \texttt{setCarma} in order to build an object of class \texttt{yuima.carma} and we show how to simulate a trajectory of the CARMA(p,q) process using the same procedure available for an object of class \texttt{yuima.model}.  
\smallskip \newline
The arguments used in a call to the constructor \texttt{setCarma()} are:

\begin{verbatim}
setCarma(p,q,loc.par=NULL,scale.par=NULL,ar.par="a",ma.par="b",lin.par=NULL,
  Carma.var="v",Latent.var="x",XinExpr=FALSE, ...)
\end{verbatim}

In the following we illustrate the arguments of the \texttt{setCarma} function:

\begin{itemize}
  \item{\texttt{p}}{ is a non-negative integer that indicates the number of the autoregressive coefficients.}
  \item{\texttt{q}}{ is a non-negative integer that indicates the order of the moving average coefficients.}
  \item{\texttt{loc.par}}{ is a string for the label of the location coefficient. The default value \texttt{loc.par=NULL} implies that $c_0=0$.}
  \item{scale.par}{ is a character-string that is the label of scale coefficient. The default value \verb|scale.par=NULL| implies that \verb|sigma=1|.}
  \item{\texttt{ar.par}}{ is a character-string that is the label of the autoregressive coefficients. The default Value is \verb|ar.par="a"|.}
  \item{\texttt{ma.par}}{ is a character-string specifying the label of the moving average coefficients. The default Value is \verb|ma.par="b"|.}
  \item{\texttt{Carma.var}}{ is a character-string that is the label of the observed process. Defaults to \verb|"v"|.}
  \item{\texttt{Latent.var}}{ is a character-string representing the label of the unobserved process. Defaults to \verb|"x"|.}
  \item{\texttt{lin.par}}{ is a character-string that is the label of the linear coefficients. If \verb|lin.par=NULL|, the default, the \texttt{setCarma} builds the CARMA(p,q) model defined as in \cite{Brockwell2001}. }
  \item{\texttt{XinExpr}}{ is a logical variable. The default value \verb|XinExpr=FALSE| implies that the starting condition for \verb|Latent.var| is zero. If \verb|XinExpr=TRUE|, each component of \texttt{Latent.var} has a parameter as a initial value.}
  \item{\texttt{...}}{ Arguments to be passed to \texttt{setCarma}, such as the slots of \verb|yuima.model-class|. They play a fondamental role when the underlying noise is a pure jump L\'evy process. In particular the following two arguments are necessary:}
  \begin{itemize}
    \item{\texttt{measure}}{ L\'evy measure of jump variables.}
    \item{\texttt{measure.type}}{ type specification for Levy measure.}
  \end{itemize}
\end{itemize}
\smallskip

Assume that we want to build a \verb|CARMA(p=3,q=1)| model driven by a standard Brownian Motion with location parameter. In this case, the state space model in \eqref{gen-state-space-mod} can be written in a explicit way as follows:
\begin{eqnarray}
Y_{t}&=& b_{0}X_{0,t}+b_{1}X_{1,t}\nonumber\\
\mbox{d}X_{0,t}&=&X_{1,t}\mbox{d}t\nonumber\\
\mbox{d}X_{1,t}&=&X_{2,t}\mbox{d}t\nonumber\\
\mbox{d}X_{2,t}&=&\left[-a_{3}X_{0,t}-a_{2}X_{1,t}-a_{1}X_{0,t}\right]\mbox{d}t+dZ_{t}\nonumber\\
\label{carma.ex1}
\end{eqnarray}
where $Z_{t}$ is a Wiener process. \newline
For this reason, we instruct \texttt{yuima} to create an object of class \texttt{yuima.carma} using the code listed below.

\begin{verbatim}
> Carma_brown_mod<-setCarma(p=3,q=1,loc.par="c0",Carma.var="y",Latent.var="X")
\end{verbatim}

We can display the internal structure of the object \verb|Carma_brown_mod using| the \textsf{R} utility \texttt{str}:

\begin{verbatim}
> str(Carma_brown_mod)
\end{verbatim}
\begin{verbatim}
Formal class 'yuima.carma' [package "yuima"] with 17 slots
  ..@ info           :Formal class 'carma.info' [package "yuima"] with 10 slots
  .. .. ..@ p         : num 3
  .. .. ..@ q         : num 1
  .. .. ..@ loc.par   : chr "c0"
  .. .. ..@ scale.par : chr(0) 
  .. .. ..@ ar.par    : chr "a"
  .. .. ..@ ma.par    : chr "b"
  .. .. ..@ lin.par   : chr(0) 
  .. .. ..@ Carma.var : chr "y"
  .. .. ..@ Latent.var: chr "X"
  .. .. ..@ XinExpr   : logi FALSE
  ..@ drift          :  expression((b0 * X1 + b1 * X2), (X1), (X2)) ...
  ..@ diffusion      :List of 4
  .. ..$ :  expression((0))
  .. ..$ :  expression((0))
  .. ..$ :  expression((0))
  .. ..$ :  expression((1))
  ..@ hurst          : num 0.5
  ..@ jump.coeff     :  expression()
  ..@ measure        : list()
  ..@ measure.type   : chr(0) 
  ..@ parameter      :Formal class 'model.parameter' [package "yuima"] with 7 slots
  .. .. ..@ all      : chr [1:8] "b0" "b1" "a3" "a2" ...
  .. .. ..@ common   : chr(0) 
  .. .. ..@ diffusion: chr(0) 
  .. .. ..@ drift    : chr [1:5] "b0" "b1" "a3" "a2" ...
  .. .. ..@ jump     : chr(0) 
  .. .. ..@ measure  : chr(0) 
  .. .. ..@ xinit    : chr [1:5] "c0" "b0" "X0" "b1" ...
  ..@ state.variable : chr [1:4] "y" "X0" "X1" "X2"
  ..@ jump.variable  : chr(0) 
  ..@ time.variable  : chr "t"
  ..@ noise.number   : int 1
  ..@ equation.number: int 4
  ..@ dimension      : int [1:6] 8 0 0 5 0 0
  ..@ solve.variable : chr [1:4] "y" "X0" "X1" "X2"
  ..@ xinit          :  expression((c0 + b0 * X0 + b1 * X1), (0), (0)) ...
  ..@ J.flag         : logi FALSE
\end{verbatim}

Looking to the structure, we observe that the slots \texttt{measure} and \texttt{measure.type} are both empty meaning that the underlying process is a standard Brownian Motion. The slots \texttt{drift} and \texttt{diffusion} contains expression that represents the CARMA(3,1) model using the following representation of system \eqref{carma.ex1}:
\begin{equation}
\mbox{d}\left[
\begin{array}{l}
Y_{t}\\
X_{0,t}\\
X_{1,t}\\
X_{2,t}\\
\end{array}
\right]=
\left[
\begin{array}{l}
b_{0} X_{1}+b_{1} X_{2}\\
X_{1,t}\\
X_{2,t}\\
-a_{3}X_{0,t}-a_{2}X_{1,t}-a_{1}X_{2,t}\\
\end{array}
\right] \mbox{d}t+
\left[
\begin{array}{l}
0\\
0\\
0\\
1\\
\end{array}
\right]\mbox{d}Z_{t}
\end{equation}
Notice that, since we define the CARMA(p,q) model using the standard \texttt{yuima} mathematical description, we need to rewrite the observable process $Y_{t}$ as a stochastic differential equation. The location parameter $c_{0}$ is contained in the slot \texttt{xinit} where the starting condition of the variable $Y_{t}$ is:
\begin{equation*}
Y_{0}=c_{0}+b_{0}X_{0} + b_{1}X_{1}
\end{equation*}
To ensure the existence of a second order solution, we choose the autoregressive coefficients $\mathbf{a}:=\left[a_{1},a_{2},a_{3}\right]$ such that the eigenvalues of the matrix $A$ are real and negative (see Prop. \ref{propsecord}). Indeed, $a_{1}=4$, $a_{2}=4.75$ and $a_{3}=1.5 $, it is easy to verify that the eigenvalues of matrix $A$ are $\lambda_{1}=-0.5$, $\lambda_{2}=-1.5$ and $\lambda_{3}=-2$. The next phase is to show the necessary steps for simulating a sample path of the model in \eqref{carma.ex1}. It is worthing to remark that, since the \texttt{yuima.carma} extends the \texttt{yuima.model}, we use the same procedure described in \cite{Brousteetal2013}. 
\smallskip

We fix the value for the model parameters: 

\begin{verbatim}
> par.Carma_brown_mod<-list(a1=4,a2=4.75,a3=1.5,b0=1,b1=0.23,c0=0)
\end{verbatim}

We set the sampling scheme:

\begin{verbatim}
> samp<-setSampling(Terminal=400, n=16000)  
\end{verbatim}

Applying the \texttt{simulate} method, we obtain an object of class \texttt{yuima} that contains the simulated trajectory:

\begin{verbatim}
> set.seed(123)
> sim.Carma_brown_mod<-simulate(Carma_brown_mod,true.parameter=par.Carma_brown_mod,
+                               sampling=samp)
\end{verbatim}

The simulated sample path can be drawn using the \texttt{plot} function. Since the simulation procedure
is based on the state space representation of the CARMA model, the \texttt{plot} function 
returns a multiple figure. The upper is the sample path of the CARMA process $Y_{t}$  while the remaining pictures report the corresponding trajectories of each component of the state vector $X_{t}$.

\begin{verbatim}
> plot(sim.Carma_brown_mod)
\end{verbatim}

\begin{figure}[h]
  \centering
		\includegraphics[width=0.7\textwidth]{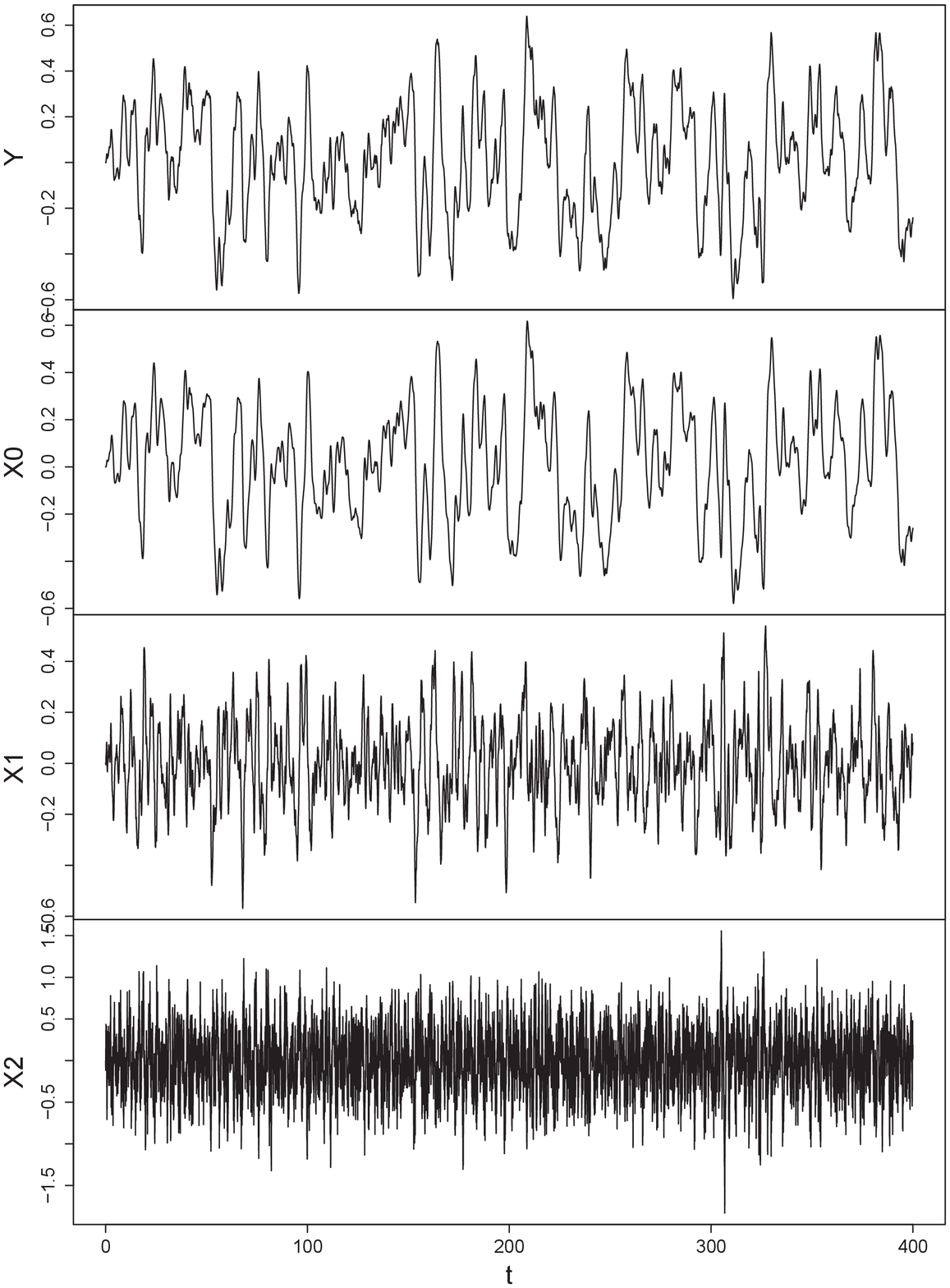}
      \caption{Simulated sample path of a CARMA(3,1) process driven by a standard Browniam motion \label{fig1} }
	\label{sim.Carma_brown_mod}
\end{figure}

Insert here figure \ref{fig1}.

\subsection{Estimation of a CARMA model}
In this Section we explain how to use the \texttt{qmle} method for performing the three steps estimation procedure described in Sect. \ref{TwoStepEst} for the CARMA(p,q) model. As reported in \cite{Brousteetal2013}, the \texttt{qmle} function implemented in \texttt{yuima} package works as similar as possible to the standard \texttt{mle} function in the \texttt{stats4} package when the model is an object of the \texttt{yuima.model-class}. However the behaviour of the function is slightly different if we considerr an object of the \texttt{yuima.carma-class}. Indeed in this case, the \texttt{qmle} function can be return an object of class \texttt{mle} or and object of class \texttt{yuima.carma.qmle-class}. \newline This class extends the existing class \texttt{mle} for the \texttt{stats4} package since it has an adjoint slot which contains the  L\'evy increments estimated by the new \texttt{yuima} function \texttt{CarmaNoise}.
\smallskip

The arguments in the function \texttt{qmle} are:

\begin{verbatim}
qmle(yuima, start, method="BFGS", fixed = list(), print=FALSE, 
  lower, upper, joint=FALSE, Est.Incr="Carma.IncPar",aggregation=TRUE ...)
\end{verbatim}

For a complete treatment of the arguments passed to the \texttt{qmle} we refer to the \texttt{yuima} documentation. In this work we focus our attention only on the character-string variable \texttt{Est.Incr} and the logical variable \texttt{aggregation}.\newline The variable \texttt{Est.Incr} manages the output of the \texttt{qmle} function. The variable \texttt{Est.Incr} assumes the following three values:
\begin{itemize}
\item{\texttt{Carma.IncPar}}{ that is the default value. In this case the function \texttt{qmle} returns an object of \texttt{yuima.carma.qmle-class} which contains the CARMA parameters obtained by quasi-maximum likelihood procedure, the estimated increments and parameters of the underlying L\'evy process. If the CARMA(p,q) model is driven by a standard Browniam motion, the behaviour of the function is identically when \verb|Est.Incr="Carma.Inc"|.}
\item{\texttt{Carma.Inc}}{  The function \texttt{qmle} returns an object of \texttt{yuima.carma.qmle-class} which contains only the CARMA parameters and the estimated L\'evy increments.}
\item{\texttt{Carma.Par}}{ In this case the output is an object of \texttt{mle-class} containing the estimated CARMA parameters obtained using the quasi maximum likelihood procedure and the parameters of the L\'evy process.}
\end{itemize}
\smallskip

The logical variable \texttt{aggregation} is related to the methodology for the estimation of the L\'evy parameter. Indeed if the variable is \texttt{TRUE}, the increments are aggregated in order to obtain the increments on unitary time intervals. 
\smallskip

In order to obtain the estimated increments of the underlying L\'evy process, the \texttt{qmle} function calls internally the function \texttt{CarmaNoise}. The call is done using the following command:

\begin{verbatim}
CarmaNoise(yuima, param, data=NULL)
\end{verbatim}

where the arguments mean:

\begin{itemize}
  \item{\texttt{yuima}}{ is a \texttt{yuima} object or an object of \texttt{yuima.carma-class}.}
  \item{\texttt{param}} { is a \texttt{list} of parameters for the CARMA model.}
  \item{\texttt{data}}{ is an object of class \texttt{yuima.data-class} contains the observations available at uniformly spaced time intervals. If \texttt{data=NULL}, the default, the \texttt{CarmaRecovNoise} uses the data in an object of \texttt{yuima}.}
\end{itemize}

Using the same example in Sect. \ref{CARMA1spec}, we list below the code for estimation of the CARMA(3,1) model:

\begin{verbatim}
>  qmle.Carma_brown_mod <- qmle(sim.Carma_brown_mod,start=par.Carma_brown_mod)
\end{verbatim}
\begin{verbatim}
Starting qmle for carma ... 

 Stationarity condition is satisfied...
 Starting Estimation Increments ...

Starting Estimation parameter Noise ...
\end{verbatim}

The function by default returns an object of class \texttt{yuima.carma.qmle} and we can see the values of estimated parameters applying the utulity \texttt{summary}:

\begin{verbatim}
> summary(qmle.Carma_brown_mod)
\end{verbatim}
\begin{verbatim}
Two Stage Quasi-Maximum likelihood estimation

Call:
qmle(yuima = sim.Carma_brown_mod, start = par.Carma_brown_mod)

Coefficients:
      Estimate  Std. Error
b0 0.975548500 0.010311774
b1 0.226960222 0.002573738
a3 1.736966408 0.003086911
a2 4.964930880         NaN
a1 3.908449535 0.002654231
c0 0.005858153 0.027901534

-2 log L: -178729.2 


Number of increments: 15997

Average of increments: -0.000016

Standard Dev. of increments: 0.160054

Summary statistics for increments:
      Min.    1st Qu.     Median       Mean    3rd Qu.       Max. 
-0.6159000 -0.1064000 -0.0013330 -0.0000156  0.1093000  0.6150000 
\end{verbatim}

\begin{figure}
  \centering
  	\includegraphics[width=0.4\textwidth, angle=270]{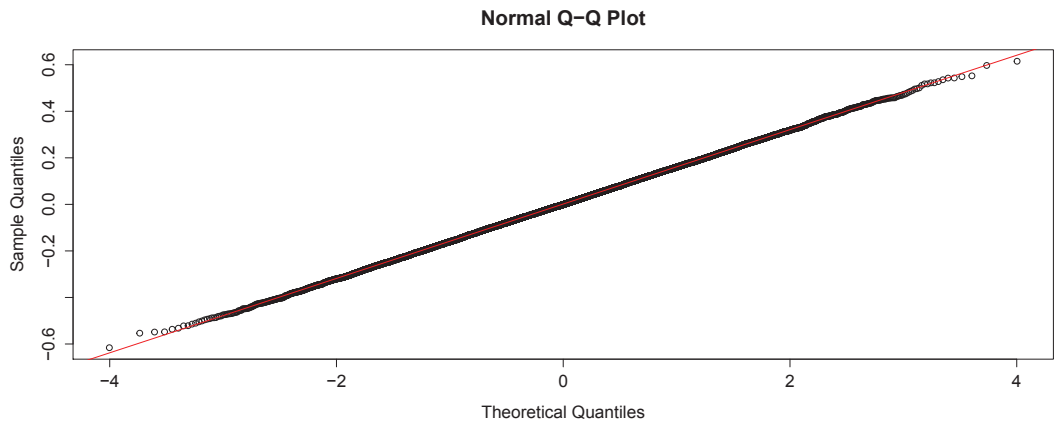}
      \caption{\texttt{QQ-norm} of the estimated increments of the underlying L\'evy process.\label{QQ_plot_Carma_brown_mod}}
\end{figure}

Since the driven noise is a standard brownian motion, then the estimated parameters are only the autoregressive
and moving average parameters. In figure \ref{QQ_plot_Carma_brown_mod}  we check the normality from a qualitative point of view using the \texttt{QQ-norm}.
\bigskip

Insert here figure \ref{QQ_plot_Carma_brown_mod}.
\bigskip

The behaviour of the \texttt{QQ-norm} seems to confirm that the estimated increments are generated from a normal distribution.

\subsection{ctarma package}

We conclude this Section by comparing the procedures illustrated before with the corresponding ones avaliable in the \texttt{ctarma} package. 
As shown in the introduction, the \texttt{ctarma} package developed by \cite{ctarma} contains several routines for the simulation and  the estimation of a Gaussian CARMA(p,q) model using both frequency and time-domain approaches. Since in this paper we focus on the state-space representation of a CARMA(p,q) model, we conduct our comparison considering only the time-domain approach and refer to \cite{Tomasson2011} for a complete and detailed explanation of the frequency-domain approach for the simulation and the estimation of a Gaussian CARMA(p,q) model.
\newline Our comparison is based on two exercises. In the first, we build an object of 
class \texttt{yuima} that contains a simulated sample path of a Gaussian CARMA(2,1) model. We write a simple function that converts an object of class \texttt{yuima} into an object of class \texttt{ctarma} and use this object for the estimation of the CARMA(2,1) parameters applying the \texttt{ctarma} function \texttt{ctarma.maxlik} that performs a maximum likelihood estimation procedure based on the Kalman Filter. We compare these results with those obtained using the \texttt{qmle} function.
\newline In the second exercise we repeat a similar experiment but in this case we simulate a trajectory of a Gaussian CARMA(2,1) using the \texttt{ctarma} function \texttt{carma.sim.timedomain}.

As first step, we simulate a trajectory of a CARMA(2,1) model using the following \texttt{yuima} functions

\begin{verbatim}
> mod.yuima<-setCarma(p=2,q=1,scale.par="sig",Carma.var="y")
> param.yuima<-list(a1=1.39631,a2=0.05029,b0=1,b1=1,sig=1)
> samp.yuima<-setSampling(Terminal=100,n=200)
> set.seed(123)
> sim.yuima<-simulate(mod.yuima,true.parameter=param.yuima,sampling=samp.yuima)
\end{verbatim}

We estimate the parameters using the \texttt{qmle} function.

\begin{verbatim}
> carmaopt.yuima<-qmle(sim.yuima,start=param.yuima)
\end{verbatim}
\begin{verbatim}
Starting qmle for carma ... 

 Stationarity condition is satisfied...
 Starting Estimation Increments ...

Starting Estimation parameter Noise ...
\end{verbatim}
\begin{verbatim}
> summary(carmaopt.yuima) 
\end{verbatim}
\begin{verbatim}
Two Stage Quasi-Maximum likelihood estimation

Call:
qmle(yuima = sim.yuima, start = param.yuima)

Coefficients:
     Estimate Std. Error
sig 2.2114666  1.9361348
b0  1.0000000  0.0000000
b1  0.5493488  0.4138758
a2  0.4223176  0.4298420
a1  3.3439956  2.4968459

-2 log L: 403.5426 


Number of increments: 198

Average of increments: 0.007277

Standard Dev. of increments: 0.609316

Summary statistics for increments:
     Min.   1st Qu.    Median      Mean   3rd Qu.      Max. 
-1.536000 -0.379400 -0.034040  0.007277  0.383400  1.760000 
\end{verbatim}

We write a simple function that converts an object of class \texttt{yuima} into an object of class ctarma:

\begin{verbatim}
> yuimaToctarma<-function(yuima,true.param){
+   if(("ctarma" %in% rownames(installed.packages()))==FALSE){
+     warning("You need to install ctarma package")
+     return(NULL)
+   }else{
+     require(ctarma)
+   }
+   if(!is(yuima,"yuima")){
+     warning("The model is not an object of class yuima")
+     return(NULL)
+   }
+   model<-yuima@model
+   if(!is(model,"yuima.carma")){
+     warning("The model is not an object of class yuima.carma")
+     return(NULL)
+   }
+   par.names<-names(true.param)
+   par<-as.numeric(true.param)
+   names(par)<-par.names
+   info<-model@info
+   p<-info@p
+   if(length(info@loc.par)!=0){
+     warning("It is not possible to convert a CARMA model with location parameter")
+     return(NULL)
+   }
+   name.ar<-paste(info@ar.par,c(1:p),sep="")
+   a<-true.param[name.ar]
+   q<-info@q
+   name.ma<-paste(info@ma.par,c(0:q),sep="")
+   b<-true.param[name.ma]
+   if(length(info@scale.par)==0){
+     sigma<-1
+   }else{
+     sigma<-par[info@scale.par]
+   }
+   
+   data<-yuima@data@zoo.data[[1]]
+   time<-index(data)
+   y<-coredata(data)
+   
+   ctarma.mod<-ctarma(ctarmalist(y,time,a,b,sigma))
+   return(ctarma.mod)
+ }
\end{verbatim}

Applying the function \texttt{yuimaToctarma} we obtain an object of class \texttt{ctarma}
 and estimate the model using the function \texttt{ctarma.maxlik}:

\begin{verbatim}
> ctarma.mod<-yuimaToctarma(sim.yuima,param.yuima)
> carmaopt.ctarma<-ctarma.maxlik(ctarma.mod)
> summary(carmaopt.ctarma)
\end{verbatim}

\begin{verbatim}
$coeff
               MLE   STD-MLE
AHAT_1   3.3447171 3.2148525
AHAT_2   0.4224339 0.5915801
B_0      1.0000000 0.0000000
BHAT_1   0.5492208 0.4104583
SIGMAHAT 2.2120532 0.3805180

$loglik
[1] -201.7713

$bic
[1] 424.7558
\end{verbatim}$
 
Now we simulate a trajectory using the \texttt{carma.sim.timedomain} function available in the \texttt{ctarma} package.

\begin{verbatim}
> a<-c(1.39631, 0.05029)
> b<-c(1,1)
> sigma<-1
> tt<-(1:200)/2
> set.seed(123)
> y<-carma.sim.timedomain(tt,a,b,sigma)
\end{verbatim}

We build an object of class \texttt{ctarma} and we estimate the model parameters using the following command lines:

\begin{verbatim}
> ctarma.mod1<-ctarma(ctarmalist(y,tt,a,b,sigma))
> carmaopt.ctarma1<-ctarma.maxlik(ctarma.mod1)
> summary(carmaopt.ctarma1)
\end{verbatim}

\begin{verbatim}
$coeff
               MLE    STD-MLE
AHAT_1   0.8290442 0.53687971
AHAT_2   0.0314847 0.05127615
B_0      1.0000000 0.00000000
BHAT_1   2.6821254 1.91361886
SIGMAHAT 0.3406472 0.10815623

$loglik
[1] -176.6141

$bic
[1] 374.4215
\end{verbatim}$

We build now an object of class \texttt{yuima.data} using the constructor \texttt{setData}

\begin{verbatim}
> yuima.data<-setData(zoo(x=matrix(y,length(y),mod.yuima@equation.number),order.by=tt))
\end{verbatim}

We build an object of class \texttt{yuima} using the constructor \texttt{setYuima} and we apply to it the \texttt{qmle} function in order to estimate the parameters of the model:  

\begin{verbatim}
> yuima.mod1<-setYuima(data=yuima.data, model=mod.yuima)
> carmaopt.yuima1<-qmle(yuima.mod1,start=param.yuima)
\end{verbatim}

\begin{verbatim}
Starting qmle for carma ... 

 Stationarity condition is satisfied...
 Starting Estimation Increments ...

Starting Estimation parameter Noise ...
\end{verbatim}

\begin{verbatim}
> summary(carmaopt.yuima1)
\end{verbatim}

\begin{verbatim}
Two Stage Quasi-Maximum likelihood estimation

Call:
qmle(yuima = yuima.mod1, start = param.yuima)

Coefficients:
      Estimate Std. Error
sig 0.34107952 0.23452372
b0  1.00000000 0.00000000
b1  2.67881403 1.78368917
a2  0.03153544 0.03349193
a1  0.82963964 0.38604787

-2 log L: 353.2283 

Number of increments: 197

Average of increments: 0.054225

Standard Dev. of increments: 0.633237

Summary statistics for increments:
    Min.  1st Qu.   Median     Mean  3rd Qu.     Max. 
-1.50000 -0.40790  0.02810  0.05422  0.43980  2.09000 
\end{verbatim}

In table \ref{tab:1} we summarize the results of our comparison:

\begin{table}[htbp]
  \centering
    \begin{tabular}{rrrrr}
    \hline
    \multicolumn{5}{c}{First Exercise} \\
    \hline
    Package & \multicolumn{2}{c}{yuima} & \multicolumn{2}{c}{ctarma } \\
    Param. & \multicolumn{1}{c}{Estimates} & \multicolumn{1}{c}{s.d} & \multicolumn{1}{c}{Estimates} & \multicolumn{1}{c}{s.d} \\
    $\sigma$ & 2.211 & 1.936 & 2.212 & 0.38 \\
    $b_0$    & 1.000 & Fixed & 1.000 & Fixed \\
    $b_1$    & 0.549 & 0.413 & 0.549 & 0.41 \\
    $a_2$    & 0.422 & 0.429 & 0.422 & 0.591 \\
    $a_1$    & 3.343 & 2.496 & 3.344 & 3.214 \\
    log L & \multicolumn{2}{c}{-201.726} & \multicolumn{2}{c}{-201.771} \\
    \hline
  	\multicolumn{5}{c}{Second Exercise} \\
    \hline
		Package & \multicolumn{2}{c}{yuima} & \multicolumn{2}{c}{ctarma } \\
    Param. & \multicolumn{1}{c}{Estimates} & \multicolumn{1}{c}{s.d} & \multicolumn{1}{c}{Estimates} & \multicolumn{1}{c}{s.d} \\
    $\sigma$ & 0.341 & 0.234 & 0.341 & 0.108 \\
    $b_0$    & 1.000 & Fixed & 1.000 & Fixed \\
    $b_1$    & 2.678 & 1.783 & 2.682 & 1.913 \\
    $a_2$    & 0.031 & 0.033 & 0.031 & 0.051 \\
    $a_1$    & 0.829 & 0.386 & 0.829 & 0.536 \\
    log L & \multicolumn{2}{c}{-176.614} & \multicolumn{2}{c}{176.614} \\
    \hline
    \end{tabular}%
  \caption{Comparison between estimation results obtained using \texttt{yuima} and \texttt{ctarma} packages. \label{tab:1}}%
\end{table}%

Looking at table \ref{tab:1} we observe that the estimates of parameters using the two packages are similar. The differences can be justified from fact that in the \texttt{ctarma} package the stationarity can be enforced using two different one-to-one transformations of the original parameters proposed by \cite{Belcher1994} and \cite{Pham1991} respectively while in the \texttt{yuima} there are no stationarity constraints and the stationarity is checked once the estimates are obtained. 
\newline Although both transformations in the \texttt{ctarma} allow to formulate the maximum likelihood estimation as an unconstrained optimization problem on the new variables, the choice in the \texttt{yuima} is justified by the following two reasons:
\begin{itemize}
\item The optimization problem is defined on the original autoregressive and moving avarege parameters and this is coherent with the spirit of the \texttt{Yuima} project.
\item Defining the optimization problem on the original variables allows the user to manage efficiently the possibility of having constraints on the model parameters.
\end{itemize}

\section{Simulation and estimation of a CARMA(p,q) model driven by a L\'evy process }
\label{NumEx}

In this Section we show how to simulate and estimate a CARMA(p,q) model driven by a L\'evy 
 process in the \texttt{yuima} package. Based on our knowledge, \texttt{yuima} is the first  
package available on CRAN that allows the user to manage, in a complete way, a L\'evy CARMA model. As shown in Sect. \ref{ImplemCarmaInYuima}, it is also possible to recover the increments of the underlying L\'evy and consequently the user can build on it a non-parametric L\'evy CARMA model, i.e. a model where the distribution of the increments is not specified.
\newline
In order to test the procedures implemented in \texttt{yuima} for the simulation and the estimation of a CARMA(2,1) model we consider three different exercises: 
\begin{itemize}
\item We simulate a trajectory from a CARMA(2,1) driven by a Compound Poisson process with normally distributed jumps and then we use this trajectory for the estimation procedure.
\item We repeat a similar exercise and assume that the underlying L\'evy process is a Variance Gamma model \cite{Madan1990}.      
\item In the last experiment, we assume the underlying L\'evy process to be a Normal Inverse Gaussian model \cite{Barndorff1977}.
\end{itemize}

It is worth to notice that since all the considered models can be seen as mixture of normals, the maximum likelihood estimation could be efficiently performed through an EM algorithm as that proposed in \cite{Demp1977} and used for the Compound Poisson \cite{Hinde}, the Variance Gamma \cite{Loregian2012} and the Normal Inverse Gaussian \cite{Karlis2002}. We prefer to maximize directly the log-likelihood function and the  densities are computed via Inverse Fourier Transforms. We leave the estimation procedure based on the EM algorithm for future developments of the \texttt{yuima} package.
  
\textbf{First example}: \newline We consider a CARMA(2,1) driven by a Compound Poisson where jump size is normally distributed and $\lambda$ is equal to 1.

\begin{verbatim}
> modCP<-setCarma(p=2,q=1,Carma.var="y",
+                 measure=list(intensity="Lamb",df=list("dnorm(z, mu, sig)")),
+                 measure.type="CP") 
> true.parmCP <-list(a1=1.39631,a2=0.05029,b0=1,b1=2,
+                   Lamb=1,mu=0,sig=1)
\end{verbatim}

We obtain a sample path of the model using the yuima's simulate function.

\begin{verbatim}
> samp.L<-setSampling(Terminal=200,n=4000)
> set.seed(123)
> simCP<-simulate(modCP,true.parameter=true.parmCP,sampling=samp.L)
> plot(simCP,main="CP CARMA(2,1) model",type="l")
\end{verbatim}

\begin{figure}[h]
  \centering
  	\includegraphics[width=0.7\textwidth]{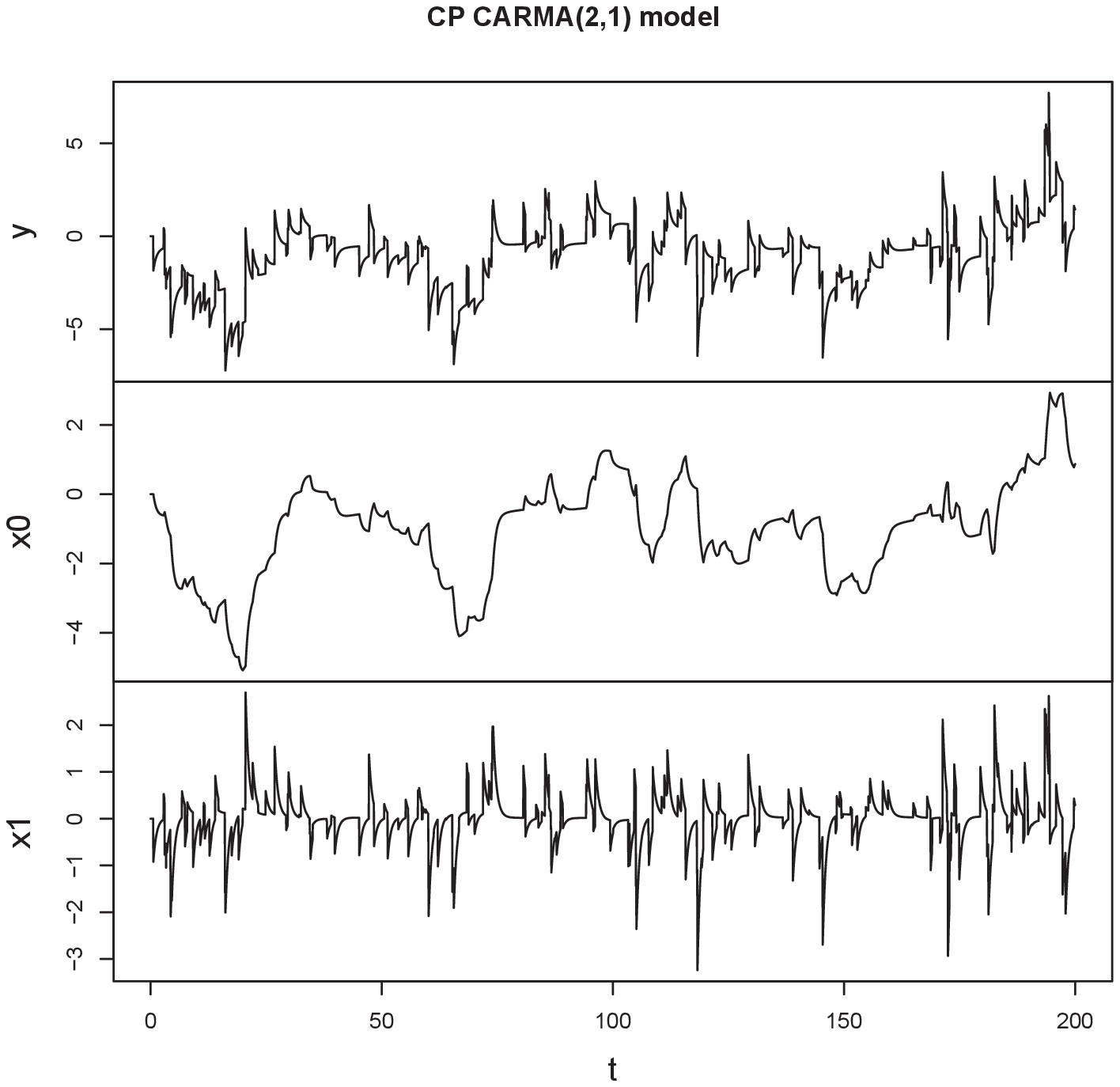}
\caption{Simulated sample path of a CARMA model driven by a Compound Poisson process. The jump size is normally distributed. \label{CompCarma}}
\end{figure}

Insert here figure \ref{CompCarma}.
\bigskip

We estimate the parameter using the three step procedure described in Sect. \ref{TwoStepEst}.

\begin{verbatim}
> carmaoptCP <- qmle(simCP, start=true.parmCP)
\end{verbatim}

\begin{verbatim}
Starting qmle for carma ... 

 Stationarity condition is satisfied...
 Starting Estimation Increments ...

Starting Estimation parameter Noise ...
\end{verbatim}

\begin{verbatim}
> summary(carmaoptCP)
\end{verbatim}

\begin{verbatim}
Two Stage Quasi-Maximum likelihood estimation

Call:
qmle(yuima = simCP, start = true.parmCP)

Coefficients:
         Estimate  Std. Error
b0    0.783877655 0.253566362
b1    1.827108561 0.021281824
a2    0.078614454 0.046810522
a1    1.384434622 0.229686130
Lamb  1.038752541 0.103628795
mu   -0.005414145 0.005461136
sig   0.984266035 0.006756150

-2 log L: 4006.412 


Number of increments: 3998

Average of increments: -0.004651

Standard Dev. of increments: 0.218475

-2 log L of increments: 432.210758

Summary statistics for increments:
      Min.    1st Qu.     Median       Mean    3rd Qu.       Max. 
-3.1580000 -0.0051220 -0.0015580 -0.0046510  0.0007187  2.7260000 
\end{verbatim}

\begin{verbatim}
> plot(carmaoptCP,main="Compound Poisson with normal jump size",ylab="Incr.",type="l")
\end{verbatim}

\begin{figure}[h]
  \centering
  	\includegraphics[width=0.4\textwidth, angle=270]{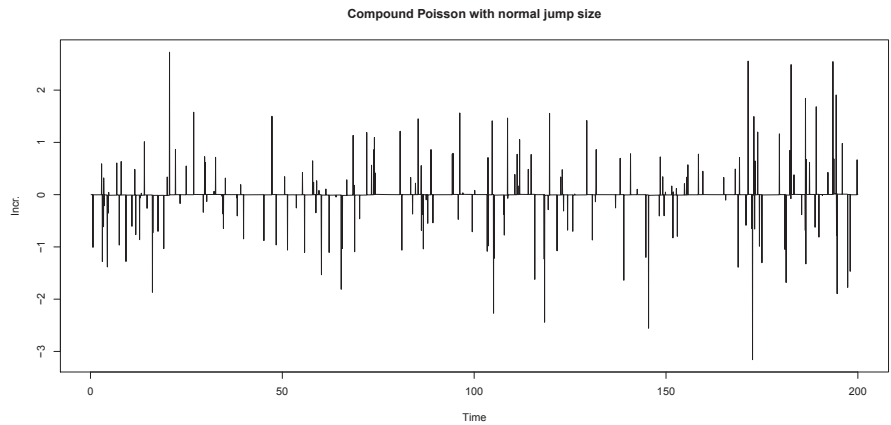}
		\caption{Estimated L\'evy Increments for a Carma model driven by a Compound Poisson process. \label{IncrCPcarma}}
\end{figure}

Insert here figure \ref{IncrCPcarma}.
\bigskip

\begin{figure}[h]
  \centering
  	\includegraphics[width=0.4\textwidth, angle=270]{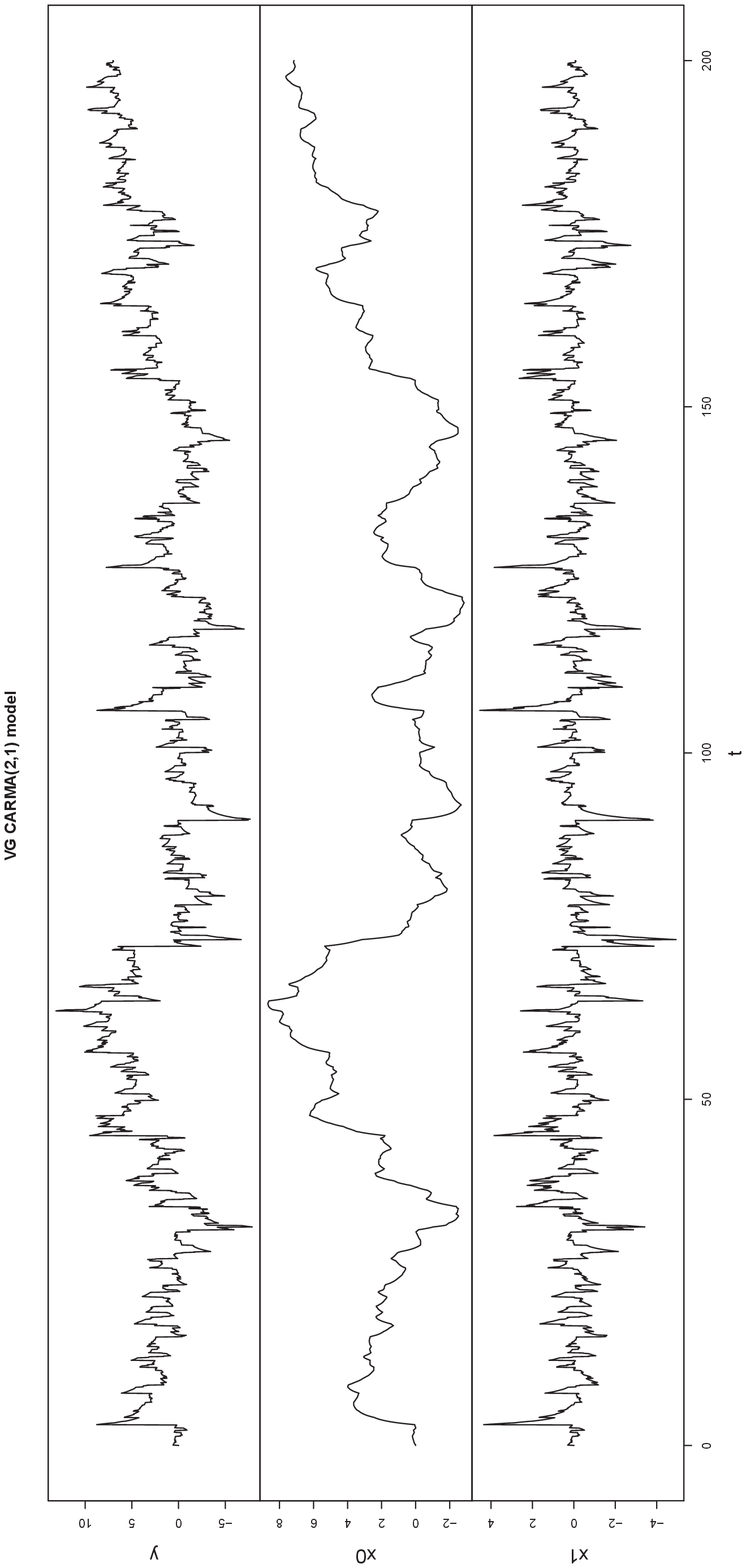}
\caption{Simulated sample path of a CARMA model driven by a Compound Poisson process. The jump size is normally distributed. \label{simVGCarma}}
\end{figure}

\textbf{Second Example}: \newline
In this case, the underlying L\'evy is a Variance Gamma model and we instruct \texttt{yuima} to build a CARMA(2,1) process with the following command line:

\begin{verbatim}
> modVG<-setCarma(p=2,q=1,Carma.var="y",
+                 measure=list("rngamma(z,lambda,alpha,beta,mu)"),measure.type="code") 
> true.parmVG <-list(a1=1.39631, a2=0.05029, b0=1, b1=2,
+                    lambda=1, alpha=1, beta=0, mu=0)
\end{verbatim}

We simulate a trajectory as follows:

\begin{verbatim}
> set.seed(100)
> simVG<-simulate(modVG, true.parameter=true.parmVG, sampling=samp.L)
> plot(simVG,main="VG CARMA(2,1) model",type="l")
\end{verbatim}

Insert here figure \ref{simVGCarma}.
\bigskip

Applying the \texttt{qmle} function we get

\begin{verbatim}
> carmaoptVG <- qmle(simVG, start=true.parmVG)
\end{verbatim}

\begin{verbatim}
Starting qmle for carma ... 

 Stationarity condition is satisfied...
 Starting Estimation Increments ...

Starting Estimation parameter Noise ...
\end{verbatim}

\begin{verbatim}
> summary(carmaoptVG)
\end{verbatim}

\begin{verbatim}
Two Stage Quasi-Maximum likelihood estimation

Call:
qmle(yuima = simVG, start = true.parmVG)

Coefficients:
          Estimate Std. Error
b0      1.39902454 0.41767662
b1      2.89637615 0.03385217
a2      0.04582997 0.03307549
a1      1.44727320 0.24192786
lambda  1.04555597 0.26067994
alpha   1.49836490 0.25063939
beta   -0.04555458 0.08291164
mu      0.14534591 0.03540655

-2 log L: 7692.227 

Number of increments: 3998

Average of increments: 0.005215

Standard Dev. of increments: 0.218392

-2 log L of increments: 524.119054

Summary statistics for increments:
      Min.    1st Qu.     Median       Mean    3rd Qu.       Max. 
-3.0330000 -0.0025170 -0.0001629  0.0052150  0.0026890  2.9550000 
\end{verbatim}

\begin{verbatim}
> plot(carmaoptVG,main="Variance Gamma increments",ylab="Incr.",xlab="Time",type="l")
\end{verbatim}

\begin{figure}[h]  
	\centering
  	\includegraphics[width=0.4\textwidth, angle=270]{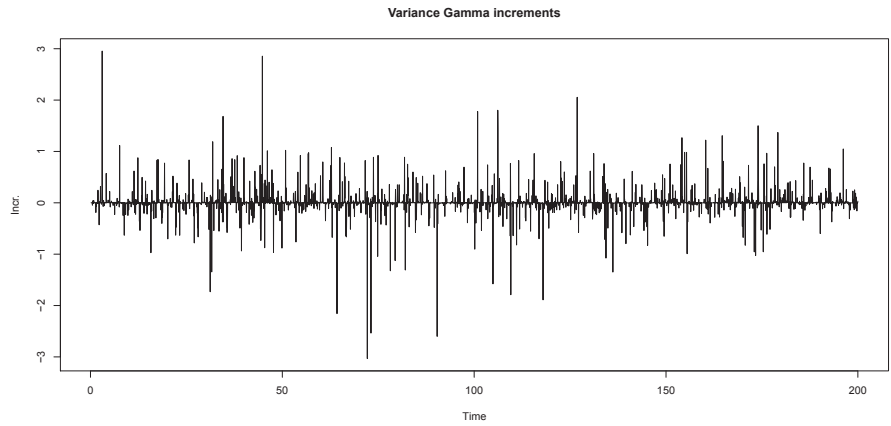}
		\caption{Estimated L\'evy increments for a Carma model driven by a Variance Gamma process \label{IncrVGcarma}}
\end{figure}

Insert here figure \ref{IncrVGcarma}.
\bigskip

\textbf{Third Example}:\newline
In the third example we assume that the underlying L\'evy is a Normal Inverse Gaussian process.
As a first step we define a CARMA(2,1) process using the yuima constructor setCarma:

\begin{verbatim}
> modNIG<-setCarma(p=2,q=1,Carma.var="y",
+                  measure=list("rNIG(z,alpha,beta,delta1,mu)"),measure.type="code") 
\end{verbatim}

In this case we build explicity the underlying L\'evy process using the \texttt{yuima} package

\begin{verbatim}
> IncMod<-setModel(drift="0",diffusion="0",jump.coeff="1",
+                  measure=list("rNIG(z,1,0,1,0)"),measure.type="code")
> set.seed(100)
> simLev<-simulate(IncMod,sampling=samp.L)
> incr.lev<-diff(as.numeric(simLev@data@zoo.data$Series))
> plot(incr.lev,main="simulated noise increments",type="l")
\end{verbatim}$

\begin{figure}[h]
  \centering
  	\includegraphics[width=0.7\textwidth]{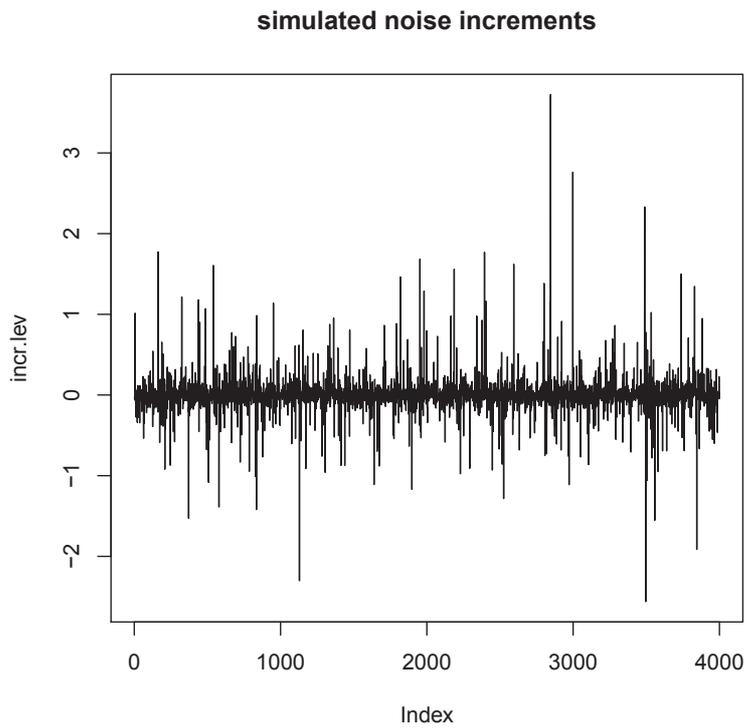}
		\caption{Simulated increments for a Normal Inverse Gaussian process \label{IncrNIG}}
\end{figure}

Insert here figure \ref{IncrNIG}.
\bigskip

%

The simulated L\'evy increments are necessary for building the sample path of the CARMA(2,1) model driven by a Normal Inverse Gaussian  process. In \texttt{yuima} package, we simulate a trajectory using the code listed below:

\begin{verbatim}
> true.parmNIG <-list(a1=1.39631,a2=0.05029,b0=1,b1=2,
+                   alpha=1,beta=0,delta1=1,mu=0)
> simNIG<-simulate(modNIG,true.parameter=true.parmNIG,sampling=samp.L)
\end{verbatim}

Applying the two steps procedure we obtain the following result:

\begin{verbatim}
> carmaoptNIG <- qmle(simNIG, start=true.parmNIG)
\end{verbatim}

\begin{verbatim}
Starting qmle for carma ... 

 Stationarity condition is satisfied...
 Starting Estimation Increments ...

Starting Estimation parameter Noise ...
\end{verbatim}

\begin{verbatim}
> summary(carmaoptNIG)
\end{verbatim}

\begin{verbatim}
Two Stage Quasi-Maximum likelihood estimation

Call:
qmle(yuima = simNIG, start = true.parmNIG)

Coefficients:
          Estimate Std. Error
b0      1.45563876 0.57914335
b1      1.96734358 0.02336632
a2      0.08624902 0.05545456
a1      1.64048955 0.41504117
alpha   1.24223088 0.39987929
beta    0.16177182 0.18718818
delta1  1.23129950 0.31792981
mu     -0.23920126 0.16707800

-2 log L: 4610.315 

Number of increments: 3998

Average of increments: -0.004164

Standard Dev. of increments: 0.218789

-2 log L of increments: 555.416356

Summary statistics for increments:
     Min.   1st Qu.    Median      Mean   3rd Qu.      Max. 
-3.138000 -0.050620 -0.001853 -0.004164  0.041450  3.283000 
\end{verbatim}

\begin{verbatim}
> plot(carmaoptNIG,main="Normal Inverse Gaussian",ylab="Incr.",type="l")
\end{verbatim}

\begin{figure}[h]
  \centering
  	\includegraphics[width=0.4\textwidth, angle=270]{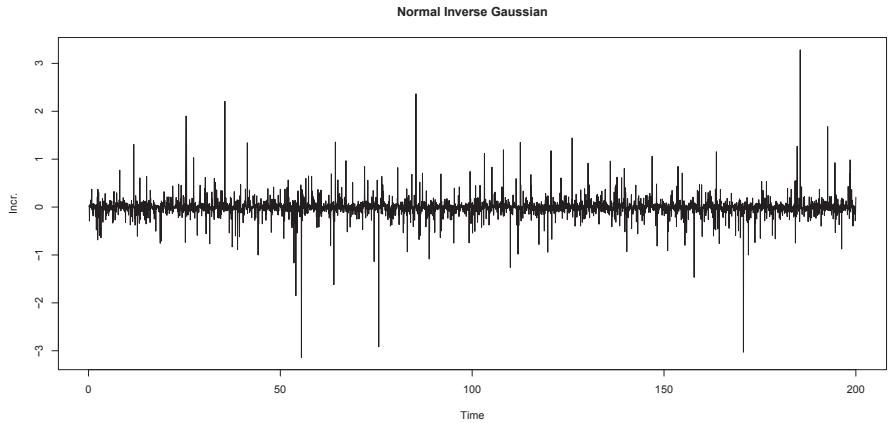}
\caption{Estimated increments for a Carma model driven by a Normal Inverse Gaussian process \label{IncrNIGCarma}}
\end{figure}

Insert here figure \ref{IncrNIGCarma}.
\bigskip
%
%

In the end of this Section, we show how to estimate the parameters of the underlying Normal Inverse Gaussian L\'evy process using the package \texttt{GeneralizedHyperbolic} and discuss the accuracy of the estimates obtained using the \texttt{qmle} function.

As a first step, we get the L\'evy shock from an object of class \texttt{yuima.carma.qmle} 

\begin{verbatim}
> NIG.Inc<-as.numeric(coredata(carmaoptNIG@Incr.Lev))
> NIG.freq<-frequency(carmaoptNIG@Incr.Lev)
\end{verbatim}

We aggregate the innovations in order to obtain the increments on the interval with unit length. 

\begin{verbatim}
> Unitary.NIG.Inc<-diff(cumsum(NIG.Inc)[seq(from=1, to=length(NIG.Inc), by=NIG.freq)])
\end{verbatim}

The function \texttt{nigFit}, available in the package \texttt{Generalized Hyperbolic}, fits
a Normal Inverse Gaussian distribution to the data \texttt{Unitary.NIG.Inc} maximizing the log-likelihood  function. The function returns an \texttt{S3} object of class \texttt{nigFit}.

\begin{verbatim}
> library(GeneralizedHyperbolic)
> FitInc.NIG.Lev<-nigFit(Unitary.NIG.Inc)
> summary(FitInc.NIG.Lev, hessian = TRUE, hessianMethod = "tsHessian")
\end{verbatim}

\begin{verbatim}
Data:      Unitary.NIG.Inc 
Hessian:  tsHessian 
               mu      delta      alpha       beta
mu    -246.191203  -9.686779   7.619589 -199.06132
delta   -9.686779 -56.595855  39.612917  -13.39664
alpha    7.619589  39.612917 -34.270069   12.82121
beta  -199.061324 -13.396643  12.821209 -191.77232
Parameter estimates:
       mu        delta       alpha        beta  
   -0.15880     1.21181     1.26833     0.08503 
  ( 0.16206)  ( 0.30642)  ( 0.39880)  ( 0.18539)
Likelihood:         -272.713 
Method:             Nelder-Mead 
Convergence code:   0 
Iterations:         247 
\end{verbatim}

Looking to the \texttt{summary}, differences on the estimates obtained with two methods, \texttt{qmle} and \texttt{nigFit}, are negligibles. In the following figure we report a comparison of the theoretical and empirical log-densities (left side) and a qqplot (right side) obtained using the \texttt{plot} function for an object of class \texttt{nigFit}.

\begin{verbatim}
> par(mfrow = c(1, 2))
> plot(FitInc.NIG.Lev, which = 2:3,
+        plotTitles = paste(c("Histogram of NIG ",
+                             "Log-Histogram of NIG ",
+                             "Q-Q Plot of NIG "), "Incr.",
+                           sep = ""))
\end{verbatim}

\begin{figure}[h]
  \centering
  	\includegraphics[width=0.4\textwidth, angle=270]{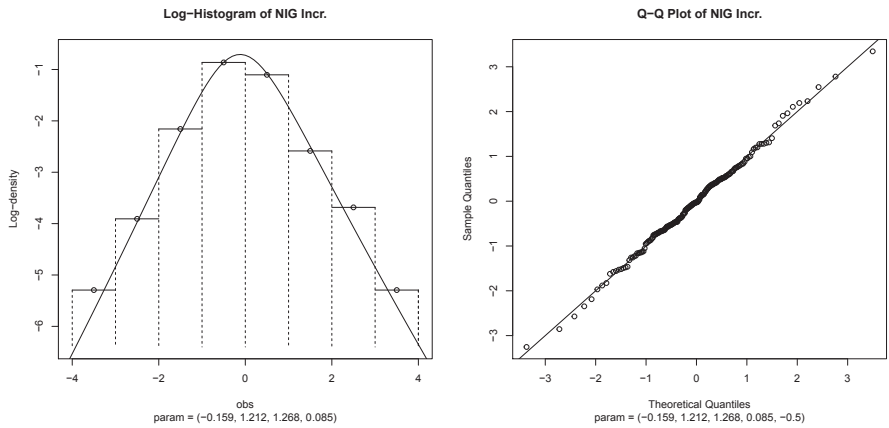}
\caption{Comparison theoretical and empirical log-density  (left side). \texttt{qq-plot} (right side) of the estimated increments. The quantiles used for comparison are computed using the Normal Inverse Gaussian distribution. \label{IncrNIG2}}
\end{figure}

Insert here figure \ref{IncrNIG2}.
\bigskip

\bibliographystyle{plain}

\end{document}